\documentclass[aps,prd,groupedaddress]{revtex4}
\usepackage[parfill]{parskip}    
\usepackage{graphicx}
\usepackage{amssymb}
\usepackage{epstopdf}
\usepackage{tcolorbox}
\usepackage{color}
\usepackage{float}
\usepackage{enumerate}
\usepackage{ulem}
\usepackage{pdfpages}
\usepackage{amsmath}
\usepackage[colorlinks=true,citecolor=blue,urlcolor=magenta,breaklinks]{hyperref}

\begin{document}

\title{Hyperbolic Casimir-like wormhole}

\author{Roberto Avalos}
\email{roberto.avalos@emory.edu}
\affiliation{Department of Physics, Emory University, Atlanta, GA 30322, USA}

\author{D. Brito}
\email{d.britou.494@gmail.com}
\affiliation{Centro de F\'isica Te\'orica y Computacional,\\ Escuela de F\'isica, Facultad de Ciencias, Universidad Central de Venezuela, Caracas 1050, Venezuela\\}

\author{E. Fuenmayor }
\email{ernesto.fuenmayor@ciencs.ucv.ve}
\affiliation{Centro de F\'isica Te\'orica y Computacional,\\ Escuela de F\'isica, Facultad de Ciencias, Universidad Central de Venezuela, Caracas 1050, Venezuela\\}

\author{E. Contreras }
\email{ernesto.contreras@gmail.com (corresponding author)}
\affiliation{Departamento de F\'{\i}sica Aplicada, Universidad de Alicante, Campus de San Vicente del Raspeig, E-03690 Alicante, Spain.\\}

\begin{abstract}
We present a systematic study of exact solutions for traversable wormhole geometries in a static and hyperbolic symmetric spacetime. In the conventional form of studying wormhole geometry, traversability requires the presence of exotic matter, which also provides negative gravity effects to keep the wormhole throat open. Using hyperbolic symmetry we obtain a solution already provided with negative energy density that replaces this effect and allows us to derive wormhole geometries that effectively violate the null energy condition. To achieve this goal, we use a generalized complexity factor for hyperbolic symmetry adapted to study wormhole geometries and with a suitable redshift function in order to construct a Casimir-like traversable hyperbolic wormhole. A detailed study has been conducted on the behavior of the matter sector, the energy conditions, and the traversability conditions.
\end{abstract}

\maketitle

\section{Introduction}

As it is well known, while the Schwarzschild solution is the only static, spherically symmetric vacuum solution, the static nature of an observer freely moving in all directions in spacetime no longer holds once the observer crosses the event horizon. This is because the event horizon also acts as a Killing horizon, causing the temporal Killing vector to become space-like after crossing it, thereby forcing the observer to inevitably end at the singularity. In this regard, no static observers can be defined inside the Schwarzschild horizon 
(see \cite{Rindler, Caroll} for a discussion on this point).  Besides, it is a well-known fact that any transformation that maintains the static form of the Schwarzschild metric (in all space-time) cannot remove the coordinate singularity in the line element \cite{Rosen}. \\

Static solutions in general relativity \cite{Allstatic0, Allstatic} are of great importance for several reasons, their mathematical simplicity makes them easier to analyze, many static solutions serve as models for astrophysical objects and studying these solutions helps us to understand properties such as the event horizon and the singularity of black holes. Furthermore, these solutions allow us to investigate the structure of spacetime in the presence of different configurations of matter and energy that affect the geometry of spacetime. If we base ourselves on the physically reasonable point of view that any equilibrium final state of a physical process should be static, it would be desirable to have a static solution over all of space-time.\\

In \cite{Herrera:2018mzq} it was proposed a way to construct a globally static spacetime by replacing the interior of the black hole with a hyperbolically symmetric submanifold which line element can be expressed as
\begin{equation}\label{hyperbolyc_empty_metric}
ds^{2}=\left(\frac{2M}{r}-1\right)dt^{2}-\frac{dr^{2}}{\frac{2M}{r}-1}-r^{2}(d\theta^{2}+\sinh^{2}\theta d\phi^{2}) \, , 
\end{equation}
where $M$ is the mass of the black hole. In this regard the whole spacetime can be thought as composed by two submanifold: spherical for $r>2M$ and hyperbolic for $r<2M$. Clearly, both regions meet only at $\theta=0$. This idea is closely related to a study published years ago by Lobo and Mimoso in \cite{lobo2010possibility}, and although it is in another context, the advantage of this symmetry is well supported in this previous work. In fact, in \cite{lobo2010possibility} the authors argue that the metric given by (\ref{hyperbolyc_empty_metric}) can be interpreted as an \textit{anti}-Schwarzschild solution, in the same sense that a de Sitter model with negative curvature corresponds to an anti-de Sitter spacetime.
A particularly interesting point raised in \cite{lobo2010possibility} is the impossibility of recovering the usual Newtonian weak-field limit for large $r$, due to a change in the spacetime signature at $r = 2M$. This lack of a Newtonian limit can be attributed to the presence of mass sources at infinity, which generate a repulsive potential. As a consequence, any test particle in the region $r < 2M$ is inevitably forced to cross the horizon.
This conclusion is reinforced in \cite{Herrera:2020bfy}, where a detailed analysis of the geodesic structure reveals that the kinematical and dynamical behavior of test particles inside the horizon is markedly different. In particular, particles are not attracted toward the center and cannot reach it regardless of their energy. The authors argue that this behavior is due to a repulsive force within the horizon, possibly associated with the gravitational vacuum.
Interestingly, \cite{Herrera:2020bfy} suggests that test particles inside the horizon may, in principle, exit this region along the axis $\theta = 0$. That is, a particle coming from infinity could cross the horizon, be repelled before reaching the center, and then escape by crossing the horizon again outward; but only along this specific axis.\\

The discussion above is sufficient to motivate the study of hyperbolic spacetimes beyond the vacuum solution represented by (\ref{hyperbolyc_empty_metric}) as was done in Ref. \cite{Hyperbolic}, where the authors studied the sources supporting this hyperbolically symmetric vacuum. Among the conclusions, they found that such source distributions can be anisotropic in pressure, with only two principal stresses being unequal and the energy density being necessarily negative. Moreover, the fluid cannot occupy the entire space within the horizon, excluding the central region. This holds true regardless of whether the energy density within the fluid distribution is regular or not. This result implies that the central region must consist of a cavity  
which aligns with the result that no test particle with finite energy can reach the center. However, it is worth mentioning that Ref. \cite{Hyperbolic} does not represent the first study on the sources of Eq. (\ref{hyperbolyc_empty_metric}). As previously noted in Ref. \cite{lobo2010possibility}, the authors found non-vacuum hyperbolic solutions that can be interpreted as traversable wormholes. Traversable wormholes were initially proposed by Flamm \cite{flamm} when establishing the Schwarzschild solution through isometric embedding. Building upon this, Einstein and Rosen \cite{Einstein} introduced a geometric structure known as the Einstein-Rosen bridge, which was later referred to as a ``wormhole'' by Wheeler \cite{Wheeler, misner}. Wormholes are tunnel-shaped geometries that connect two distinct universes or two separate regions of the same universe, and have long intrigued theoretical physicists, particularly in the context of spherical symmetry. Traversable wormholes (popularized after the work of Morris and Thorne \cite{thorne}) possess a fundamental feature: a minimum distance that defines a tunnel connecting two regions. Following their groundbreaking paper, numerous wormhole solutions have been explored. Among them, the Ellis wormhole is one of the most commonly studied solutions in General Relativity \cite{E1,E2}. Nevertheless, keeping such a tunnel open requires exotic matter that violates the null energy condition \cite{wormholes_1988_morris, morris1988wormholes}. One way to achieve this violation is via a fluid with negative energy density exceeding its pressure in magnitude. This scenario has been examined in the construction of Casimir-like traversable wormholes with spherical symmetry, where the energy density is negative by design \cite{garattini2019casimir, Visser, Lobo0, contreras_2022_construction}. In this context, the theoretical feasibility of wormholes remains an active line of research (see \cite{lobo} for a comprehensive discussion on wormholes; see also \cite{Mustafa:2022fxn,Mustafa:2023kqt,Maurya:2024jos,Kiroriwal:2024moj,Kumar:2024vko,Singh:2024hap,Chojnacki:2021xtr,Baruah:2021fge, Kundu:2021nwp,Santiago:2021xjg,Stuchlik:2021guq,Bronnikov:2021ods,Blazquez-Salcedo:2021udn,Bronnikov:2021piw,Churilova:2021tgn,Konoplya:2021hsm,Tello-Ortiz:2021kxg,Bambi:2021qfo,Sarkar:2021uob, Capozziello:2020zbx,Blazquez-Salcedo:2020czn, Berry:2020tky,Maldacena:2020sxe, Garattini:2020kqb, Bhar, Avalos:2022tqg, Annals2} for recent developments). Among these cited works we can find some deep studies on the existence and stability of wormholes in the context of dark matter and modified theories of gravity. For instance, \cite{Mustafa:2022fxn} investigates the possibility of generalized wormhole formation within galactic halos using observational data in matter-coupled gravity, while \cite{Mustafa:2023kqt} studies similar configurations in the framework of symmetric teleparallel gravity. The impact of realistic dark matter profiles, such as the Einasto and isothermal distributions, is examined in \cite{Maurya:2024jos} within $f(\mathfrak{R}, L_m)$ gravity, and in \cite{Kiroriwal:2024moj} using $f(Q)$ gravity. Meanwhile, \cite{Kumar:2024vko} analyzes the stabilizing role of dark matter on wormholes in de Rham–Gabadadze–Tolley-like massive gravity. Evolving and conservative wormhole models are developed in $f(R,T)$ gravity \cite{modificado} and generalized $\kappa(R,T)$ frameworks \cite{Singh:2024hap}, respectively. On a more formal level, \cite{Chojnacki:2021xtr} proposes a finite action principle for spacetimes with wormhole topology. Wormholes are also studied in the context of quantum gravity and observational signatures in \cite{Kundu:2021nwp}, \cite{Blazquez-Salcedo:2021udn}, and \cite{Konoplya:2021hsm}, which explore holography, fermionic-supported wormholes, and gravitational wave echoes, respectively. Finally, \cite{Bronnikov:2021piw} presents exotic wormhole solutions in general relativity with unusual asymptotics, while \cite{contreras_2022_construction} and \cite{Explorar} provide constructive examples and reviews within modern gravitational frameworks. In this sense, such a construction is naturally accommodated in the framework of hyperbolic symmetry, since, as previously discussed, the energy density is negative by construction. In other words, wormhole solutions with negative energy density emerge naturally in hyperbolic symmetry without the need to invoke additional mechanisms.\\

At this point some comments are in order. First, as pointed out in \cite{lobo2010possibility}, the spatial part of the metric~(\ref{hyperbolyc_empty_metric}) can be associated with the hyperboloid
\begin{equation}
w^{2} + x^{2} + y^{2} - z^{2} = \left( \frac{\beta^{2}}{r^{2}} - 1 \right) r^2,
\end{equation}
(where $\beta$ will play de the role of shape function of the wormhole as explained below), embedded in a four-dimensional flat space. This correspondence suggests that hyperbolic symmetry provides a natural geometric framework for constructing the type of traversable wormhole geometry under consideration. Indeed, as we shall discuss later, in traversable wormhole configurations one has $\beta > r$, and the throat is located at the radius $r_0$ where $\beta(r_0) = r_0$. At this point, the right-hand side of the hyperboloid equation vanishes, indicating that the spatial geometry reaches its minimal radius, that is, the wormhole throat corresponds to the minimal surface of the embedded hyperboloid. Second, in \cite{Hyperbolic}, it is suggested that the inner cavity of the fluid may be filled with an additional source lacking hyperbolic symmetry. However, for the purpose of constructing a traversable wormhole, we assume that the region inside the cavity does not belong to the physical spacetime. Instead, the minimum radius, defined by the condition $\beta(r_0) = r_0$, corresponds to the throat of the wormhole, where two universes are joined by the tunnel. In this context, the full spacetime is regarded as the union of three submanifolds, $\mathcal{M}_{\text{fluid}} \cup \mathcal{M}_{\text{h-vacuum}} \cup \mathcal{M}_{\text{s-vacuum}}$, where $\mathcal{M}_{\text{fluid}}$ denotes the region filled with the matter source under consideration, $\mathcal{M}_{\text{h-vacuum}}$ corresponds to the hyperbolic vacuum region supporting the wormhole geometry, and $\mathcal{M}_{\text{s-vacuum}}$ represents the asymptotic Schwarzschild vacuum. Finally, it is worth noting that the authors in \cite{lobo2010possibility} found that their solutions are not asymptotically flat, reflecting the absence of an asymptotic region. Although not explicitly stated, their solutions can be interpreted as residing within a Schwarzschild black hole, as formalized in \cite{Herrera:2018mzq}. This is precisely the viewpoint adopted in the present work: we assume that the wormhole geometry is located inside the Schwarzschild black hole, similar to how certain fluid configurations with hyperbolic symmetry occupy this region, as shown in \cite{Hyperbolic}.\\

The  approach we follow in this work clearly differs from that in \cite{lobo2010possibility}, since here we must match the traversable wormhole geometry to a surface $r_{\Sigma} < 2M$, from which the hyperbolic vacuum (\ref{hyperbolyc_empty_metric}) becomes valid. This strategy has been employed in spherically symmetric spacetimes and typically leads to the presence of thin shells, due to the discontinuity in the second fundamental form at $r_{\Sigma}$.  The main objective of this work is to construct hyperbolic wormholes without thin shells, ensuring that the Darmois conditions are satisfied. From a technical standpoint, this leads to a system of three differential equations with five unknown functions, which requires the specification of two additional conditions. In this regard, we follow a strategy previously applied in the spherically symmetric case, which consists of prescribing the redshift function and the complexity factor of a generalization of a Casimir-like wormhole \cite{Avalos:2022tqg}.
\\

This work is organized as follows. In the next section, we review the main aspects of fluid distributions in hyperbolic symmetry developed in \cite{Hyperbolic} in addition to briefly introducing the complexity factor \cite{Complex} to be used as an additional condition for the solution. In section \ref{traversable} we review the basic aspects of traversable wormholes adapted to the previous considerations of the hyperbolic symmetry introduced. Then, in section \ref{TWC} we explore how to construct a Casimir-like traversable hyperbolic wormhole imposing certain non--vanishing complexity. The last section is dedicated to the conclusions and final remarks of the work.

\section{Field Equations}

We consider hyperbolically symmetric distributions of static fluid, locally anisotropic and which may be bounded by a surface $\Sigma$ whose
equation is $r = r_{\Sigma}$ = constant (which is the outer boundary of $\mathcal{M}_{fluid}$). Here we propose the physical variables and the basic set of equations required to describe this hyperbolically symmetric, static and locally anisotropic matter distribution. The use of a hyperbolic symmetry, rather than the conventional spherical symmetry, is physically motivated by the desire to explore configurations where the spatial curvature naturally supports negative energy densities. This may have relevance in semiclassical contexts, where quantum effects can induce effective exotic matter. Moreover, locally anisotropic fluids are not just a mathematical generalization, they often emerge in models of highly compact or self-interacting systems, including exotic stars or modified dark sector theories. The metric, in polar coordinates, is given by
\begin{align}
ds^2 = e^\nu dt^2 -e^\lambda dr^2 -r^2(d\theta^2 + \sinh^2{\theta}d\phi^2), \label{lem}
\end{align}
where, due to the assumed symmetry, $\nu(r)$ and $\lambda(r)$ are functions of $r$ only and we number the coordinates: $x^{0}= t$; $x^{1}= r$; $x^{2}= \theta$; $x^{3}= \phi$.  
The metric (\ref{lem}) has to satisfy the Einstein field equations \footnote{In this work we assume units such that $c=G=1$ and as a consequence $\kappa^{2}=8\pi$}
\begin{align}
G_{\mu \nu} = R_{\mu \nu} -\frac{1}{2}Rg_{\mu \nu} =\kappa^{2}T_{\mu \nu}. \label{Eq}
\end{align}
The most general canonical algebraic decomposition of
a second order symmetric energy-momentum tensor (compatible
with staticity and axial symmetry) satisfying our assumptions, is given by
\begin{align}
T_{\mu \nu} = (\rho + P_\perp) V_\mu V_\nu -P_\perp g_{\mu \nu} + (P_r -P_\perp) K_\mu K_\nu, \label{emt}
\end{align}
where $\rho$, $P_r$ and $P_\perp$ are the energy density, the radial pressure and the tangential pressure, respectively. Due to the symmetry under consideration, the physical variables can only denpend on $r$. On the other hand, $V_\mu$ are the four velocity components, which in our case is given by
\begin{align}
V_\mu = e^{\nu/2}\delta_\mu^0,
\end{align}
and 
\begin{align}
K_\mu =  -e^{\lambda/2}\delta_\mu^1,
\end{align}
together with the vectors  
\begin{align}
L_\mu = -r\delta^2_\mu, \quad S_\mu = -r\sinh{\theta}\delta^3_\mu, 
\end{align}
can define a canonical orthonormal tetrad. Introducing (\ref{lem}) and (\ref{emt}) in the system of field equations (\ref{Eq}), it can be shown that the only non-vanishing components of Einstein equations are given by
\begin{align}
8\pi \rho &=  - \frac{(e^{-\lambda}+1)}{r^2}+\frac{\lambda'}{r}e^{-\lambda}, \label{ec1} \\
8\pi P_r &=   \frac{(e^{-\lambda}+1)}{r^2} + \frac{\nu'}{r}e^{-\lambda},  \label{ec2}\\
8\pi P_\perp & =  \frac{e^{-\lambda}}{2}\left(\nu'' + \frac{\nu'^2}{2}-\frac{\lambda'\nu'}{2}+\frac{\nu'}{r}-\frac{\lambda'}{r}\right), \label{ec3}
\end{align}
where primes indicate derivatives respect to $r$. It is worth stressing the differences between these equations and the corresponding to the spherically symmetric
case (see for example \cite{Leon, Lane-Emden}). From these equations or using the conservation laws $T^{\mu}_{\nu;\mu} = 0$, we can get the generalized Tolman-Oppenheimer-Volkoff hydrostatic equilibrium equation for anisotropic matter \textcolor{blue}{\cite{Oppenheimer:1939ne}}
\begin{align}
P'_r + (P_r + \rho)\frac{\nu'}{2}+\frac{2}{r}\Pi = 0, \label{TOV}
\end{align}
where $\Pi =-\Delta= P_r -P_\perp$.

Following the results presented in \cite{Hyperbolic}, we shall define the mass function as 
\begin{align}
m(r) = -\left(\frac{r}{2}\right) R^3_{232} = \frac{r(1+e^{-\lambda})}{2}, \label{m1}
\end{align}
where the Riemann tensor component $R^{3}_{232}$, has been calculated with (\ref{lem}). Thus, using (\ref{ec1}), we can write
\begin{align} 
m'(r) = -4\pi r^2 \rho. \label{m2}
\end{align}

Now, using the mean value theorem to evaluate the mass integral the regularity at the center of the distribution implies that the mass function must vanish as $r^3$. However, from  (\ref{m1}) it is clear that this is not our case. Thus, the hyperbolically symmetric fluid cannot fill the space surrounding the center. Thus, as discussed in \cite{Hyperbolic}, there should be a cavity either empty, or filled with a fluid not endowed with hyperbolic symmetry. The situation just described is fully consistent with the results obtained in \cite{Herrera:2020bfy} where it was shown that test particles cannot reach the center for any finite value of its energy.

On the other hand, from (\ref{m2}), it is clear that to get a positive mass function we need $\rho \leq 0$. Thus, there is a violation of the week energy condition. Furthermore, we can write the mass function as
\begin{align}\label{m3}
m = 4\pi \int_{r_{min}}^r \!\!\! |\rho|r^2dr,
\end{align}
where due to the fact that $\rho$ is negative, we have replaced it by $-|\rho|$ (as we shall do from now on). Next, using (\ref{ec2}) and (\ref{m1}) we obtain
\begin{align}\label{nuprima}
\frac{\nu '}{2}=\frac{4\pi r^3 P_r - m}{r(2m-r)},
\end{align}
from which we may write (\ref{TOV}) as
\begin{align}
P'_r + (P_r-|\rho|)\frac{4\pi r^3 P_r - m}{r(2m-r)}+\frac{2}{r}\Pi = 0, \label{TOV2}
\end{align}
where $\Pi=P_r -P_{\perp}$. This represents the hydrostatic equilibrium equation for our fluid. The first term in (\ref{TOV2}) is just the pressure gradient (usually negative and opposing gravity). The second term describes the gravitational ``force'' and contains two different contributions. The first one is the ``passive gravitational'' mass density $P_{r} - |\rho|$, which we expect to be negative. Also the term $4\pi r^{3} P_{r} - m$ (composed by the self–regenerative pressure effect $4\pi r^{3} P_r$ minus the mass function) that is proportional to the ``active gravitational mass'', and which is negative if $4\pi r^{3} P_r < m$ (Let us recall that in general relativity, the active gravitational mass is associated with the source of the gravitational field, that is, the component of the energy-momentum tensor that enters the Einstein field equations. The passive gravitational mass, on the other hand, refers to how an object responds to a given gravitational field. Bondi showed that in a relativistic context, these concepts, along with the inertial mass, must be carefully distinguished but can coincide under specific assumptions. In our context, the active gravitational mass determines the structure of the wormhole spacetime, while the passive mass would affect how test particles (or travelers) move within it \cite{Bondi:1957zz}). As a consequence, the whole second term is positive (as expected). However, because of the equivalence principle, we must keep in mind what a negative passive gravitational mass means in the hydrostatic description compared to their usual roles with respect to the positive energy density case. Finally, the third term describes the effect of the pressure anisotropy. 

Just as we have mentioned, we assume that the fluid is bounded by a hypersurface $\Sigma$. Moreover, we will suppose that the space for $r >r_{\Sigma}$ is described by the line element (\ref{hyperbolyc_empty_metric}), 
thus the matching conditions lead to 
\begin{align}\label{emc}
e^{\nu_{\Sigma}}  &= \frac{2M}{r_{\Sigma}}-1, \quad e^{-\lambda_{\Sigma}} = \frac{2M}{r_{\Sigma}}-1, \quad P_r(r_{\Sigma})=0. \nonumber \\ &
\end{align}

We emphasize that the focus of this work is the study of traversable wormholes. In this context, the throat of the wormhole is located at the radius corresponding to the inner cavity discussed in \cite{Hyperbolic}. We do not assume that the cavity is filled with any additional matter content; rather, the points within the cavity are not considered part of the manifold on which our geometry is defined.
\\

In order to solve the system of equations (\ref{ec1}-\ref{ec3}), it is necessary to provide additional information. Specifically, we must introduce two equations that relate the metric functions and/or the matter sector to reduce the number of degrees of freedom from five to three. As we will discuss in the next section, the proposed approach involves specifying the $g_{tt}$ component of the metric and the complexity factor, which, for a hyperbolically symmetric fluid, is given by \cite{Complex}:  
\begin{align}\label{complex}
    Y_{TF} =  \frac{4\pi}{r^3} \int^r_{0} \underline{r}^3 |\rho'| d\underline{r}  - 8 \pi \Pi \;.
\end{align}  
$Y_{TF}$ encompasses the influence of local pressure anisotropy and density inhomogeneity on Tolman mass, and describes how these two factors modify the value of Tolman mass, with respect to its value for a homogeneous isotropic fluid. This definition is based on the intuitive idea that the least complex gravitational system should be characterized by a homogeneous energy density distribution with isotropic pressure. This fact was at the origin of the definition of complexity provided in \cite{Complex}.

It is worth emphasizing that a similar strategy was implemented in \cite{Avalos:2022tqg} to analyze spherically symmetric wormholes with Casimir-like complexity in spherically symmetric spacetimes. It was provided a particular value for $Y_{TF}$ that allows us to define a kind of equivalence class of solutions; namely, two solutions with the same complexity factor are equivalent. In that work, the lapse function was taken as a generalization of the one proposed in \cite{garattini2019casimir} for a Casimir wormhole. Furthermore, a modified complexity factor was introduced, accounting for the fact that the minimum radius is $r_0$ rather than $r=0$, as is the case for spherical fluids.

\section{Traversable Wormholes with Hyperbolic Symmetry} \label{traversable}
In this section, we summarize the main properties that a solution of the Einstein field equations must satisfy
to describe a traversable wormhole with hyperbolic symmetry. Before delving into the mathematical construction, it is important to recall the physical motivations behind studying wormholes with hyperbolic symmetry. In addition to being mathematically rich, such solutions may arise in scenarios involving anti-de Sitter backgrounds or topologically nontrivial spacetime regions. This opens the door for exploring whether such configurations might emerge in semiclassical gravity or modified theories with quantum corrections.

We start from the line element similar to the one proposed in \cite{lobo2010possibility}: 
\begin{align}\label{lobo_metric}
    ds^2 &= -e^{2\alpha} dt^2 + \frac{dr^2}{\frac{\beta}{r} - 1} + r^2 d\theta^2 + \left( r \sinh{\theta} \right)^2 d\phi^2  \;,
\end{align}
where $\alpha$ and $\beta$ correspond to the lapse (redshift) and shape function respectively, both being functions of the radial coordinate only. At this point, a couple of comments are in order. First, the signature differs from the one used in the previous section where we introduced the field equations for fluid distribution. Second, $0<\theta<\pi$ in contrast to $-\infty<u<\infty$ used in \cite{lobo2010possibility}. 
\\

To evaluate the Einstein equations, we consider an anisotropic source compatible with hyperbolic symmetry, that is, $T_\mu^{\, \nu} = \text{diag}(-\rho, P_r, P_\perp , P_\perp)$. In this case, the field equations yield the following system:
\begin{align}
8 \pi \rho & = -\frac{\beta'}{r^2} \; , \label{h_density_eq} \\
8 \pi P_r  & =\frac{2 \alpha '}{r}\left( \frac{\beta}{r} - 1 \right) +\frac{\beta}{r^3}  \; , \label{h_pressureR_eq}	 \\   
8 \pi P_\perp & =  \left(  \frac{\beta}{r} -1 \right)\left( \alpha ''  + (\alpha ' )^2 \right) + \frac{\alpha' \! \left( r \beta' + \beta - 2r \right)}{2r^2}\nonumber\\ &+\frac{\left( r\beta' - \beta \right)}{2r^3} \; .	\label{h_pressureP_eq} 
\end{align}
As there is no horizon, $g_{tt}$ must be a non vanishing function to avoid the existence of an infinite redshift surface, then 
$\alpha$ must be finite everywhere.\\

The information about the throat of the wormhole is encoded in its shape function $\beta$. The embedding function is constructed analogously to the spherical case. Following the ideas in \cite{lobo2010possibility}, we take a spatial slice where \( \sinh{\theta} = 1 \) without significant loss of generality and a fixed time, $t=constant$, reducing the line element to 
\begin{align}
    ds^2 &=  \frac{dr^2}{\frac{\beta}{r} - 1} + r^2 d\phi^2 \; .
\end{align}
From the embedding on a cylindrical surface $ds^2 = dr^2 + r^2 d\phi^2 + dz^2$, we obtain the relation:
\begin{align}\label{H_embeding}
    \frac{dz}{dr} = \pm \sqrt{\frac{2r- \beta}{\beta - r}} \; ,
\end{align}
where $z$ is the embedding function. As in the spherical case, the wormhole throat has a constant minimum value \( r_0 \), where the shape function also takes this value, i.e., \( \beta(r_0) = r_0 \). An important difference from the spherical case is that the shape function is bounded, namely 
\begin{align}\label{shape_cote}
    r_0 \leq \beta \leq 2 r \;.
\end{align}

The smoothness of the geometry throughout the tunnel is desirable, so we impose that the inverse of the embedding function satisfies the flaring-out condition. To be more precise, as $dr/dz=0$ (a minimum) at the throat, we impose:
\begin{align}
    \frac{d^2r}{dz^2} > 0 \; \Longrightarrow \, \frac{1}{2} \frac{\beta'r - \beta}{(2r -\beta)^2} > 0 \; .
\end{align}
Note that at the throat we obtain 
\begin{align}
\beta'(r_0) > 1.
\end{align}
Then, by using (\ref{h_density_eq}) and
(\ref{h_pressureR_eq}) we get
\begin{align}
    8\pi \Big(\rho + P_r\Big) &= \frac{1}{r^2} \left( \frac{\beta}{r} - \beta' \right) + \frac{2 \alpha '}{r^2}(\beta - r) \; ,
\end{align}
and shows that at the throat,
\begin{align}\label{null_energy_condition_hw}
    \left. 8\pi \Big(\rho + P_r\Big) \right|_{r = r_0} & =  \frac{1-\beta'(r_0)}{r_0^2} \; < \; 0 \; ,
\end{align}
the null energy condition is violated. Here, the flaring-out condition and energy density given by \eqref{h_density_eq} imply that at the throat the energy density is negative, violating both the weak and the null energy conditions. A notable difference from the case of spherical symmetry is that in that case, the null-energy condition is violated, but the energy density does not necessarily have to be negative. From a physical perspective, the violation of the energy conditions in this context does not necessarily imply that the solution is unphysical. Instead, it may reflect effective behaviors associated with semiclassical effects or modified gravity regimes. In particular, the appearance of negative energy densities under hyperbolic symmetry may indicate that such geometries naturally accommodate exotic features without requiring explicitly exotic sources. This could open new avenues for connecting wormhole configurations to physically realistic scenarios in strong gravity or early-universe models.
\\

Finally, we can obtain an expression of the complexity factor adapted to the metric potentials of the wormhole, so in the context of traversable hyperbolic wormholes, the complexity factor (\ref{complex}) reads
\begin{align} \label{complex2}
Y_{TF} &=\left(\frac{\beta}{r} - 1 \right) \left( \alpha'' + (\alpha')^2 \right) + \frac{\alpha'}{r} \left( \frac{\beta'}{2} - \frac{3\beta}{2 r} + 1 \right)\nonumber\\
&+ \frac{1}{2r^3} \left( r_0 \beta'(r_0) - 3\beta(r_0) \right),
\end{align}
from where it is straightforward to observe that a traversable wormhole, with a constant redshift function, fulfills the vanishing complexity condition in a trivial way. It is worth mentioning that Eq. (\ref{complex2}) is the first expression for the complexity factor of a traversable wormhole. Next, we may provide a particular value of $Y_{TF}$ and then use this information to find a family of solutions with the same complexity factor \cite{Avalos:2022tqg}. In this case, particular values of $Y_{TF}$ allow one to define a kind of equivalence class of solutions; namely, two solutions with the same complexity factor are equivalent.\\

 At this stage, two different approaches can be considered. One can assume that the manifold is infinite and has hyperbolic symmetry. In this case, the solution can be required to be asymptotically flat, that is, we impose conditions such that when $r \to \infty$ the metric, in that limit \eqref{lobo_metric}, leads to $ds^{2}= - dt^{2}-dr^{2}+r^{2} d \theta^{2}+r^{2} \sinh^{2} \theta d\phi^{2}$, whose Riemann tensor is zero, so the solution has a pseudo-Euclidean behavior at infinity. We find this aspect problematic because, when considering slices with constant $\theta$ and $\phi$, radial geodesics are not allowed due to the absence of the light cone structure. The other assumption that can be made is that the wormhole solution is an interior solution to the horizon. In that case, Darmois conditions must be applied to join the wormhole solution \eqref{lobo_metric} with the vacuum solution in hyperbolic symmetry \eqref{hyperbolyc_empty_metric}. In this case, the wormhole solution is located in a region within the event horizon, where the symmetry is hyperbolic and the spacetime description is given by \eqref{hyperbolyc_empty_metric}. Within the horizon, the wormhole solution is connected to the vacuum solution, so there is a value \( r_\Sigma \) between \( r_0 \) and \( r = 2M \) where the Darmois conditions are satisfied. The latter is the case we consider in this work.\\


\section{Casimir-like Hyperbolic traversable wormhole}
\label{TWC}
As we stated previously, constructing a traversable wormhole geometry from the Einstein field equations requires the introduction of two supplementary conditions to close the system. In this regard, the most common approach is to provide an equation of state and a specific choice for either one of the metric potentials (the redshift or the shape function), or for a particular component of the energy-momentum tensor. A subtle but important point is that the choice must not only make the system solvable, but also be well-motivated. This is especially relevant given that the construction of traversable wormholes involves exotic matter, and it becomes necessary to propose a plausible physical medium to account for it. In this context, we propose a wormhole sustained by Casimir-like matter, not only as a mathematically tractable model but also because the Casimir effect is one of the few known mechanisms capable of producing negative energy densities in a controlled and experimentally accessible manner. Unlike previous constructions where exotic matter extends over the entire spacetime, we aim to confine the exotic sector to a compact region inside a hyperbolic cavity, matched smoothly to a vacuum exterior. This makes our model potentially more compatible with physical expectations and semiclassical scenarios where negative energy is localized. The Casimir effect, first predicted by H. B. G. Casimir in 1948 \cite{Casimir:1948dh}, arises from quantum vacuum fluctuations between two perfectly conducting parallel plates, which result in a measurable attractive force. This phenomenon provides a rare example of macroscopic quantum effects with no classical counterpart. Due to the negative energy densities associated with the vacuum fluctuations, the Casimir effect has been widely proposed as a physically plausible source of exotic matter capable of supporting traversable wormholes. The theoretical prediction was experimentally confirmed in 1997 by S. K. Lamoreaux \cite{Lamoreaux:1996wh}, who measured the force between a conducting sphere and a flat plate at micrometer separations, in agreement with Casimir's original theory. This experimental validation solidifies the Casimir effect as a credible candidate for generating the negative energy required in wormhole physics.

In \cite{garattini2019casimir}, the author constructed a traversable Casimir wormhole in a spherically symmetric spacetime based on the following prescription
\begin{align}
    \rho \;=\;  - \frac{\hbar c \pi^2}{720 r^4} \; = \; - \frac{\kappa}{8\pi}\frac{1}{r^4} \; \; \; \; & , & \; \; \kappa = \frac{\hbar c \pi^3}{90}   \nonumber \\
    \displaystyle P_r \;=\; \omega \rho     \; \;  &,& \; \; \omega = \text{const.}    \label{EoS_hyperbolic} 
\end{align}
This result was then used as the starting point in \cite{Avalos:2022tqg} to construct traversable wormholes with Casimir-like complexity, namely, wormhole geometries whose complexity factor generalizes the one associated with the solution reported in \cite{garattini2019casimir}. In the present work, we follow the same strategy givin in \cite{Avalos:2022tqg} to construct a Casimir-like wormhole in hyperbolic symmetry. \\

First, let us start by replacing the matter sector given by (\ref{EoS_hyperbolic}) in the field equations (\ref{h_density_eq}-\ref{h_pressureP_eq}). After doing so we obtain
\begin{align}
    \displaystyle \alpha&= \frac{\omega-1}{2}\log{\left| \frac{r \omega}{r \omega + r_0} \right|} \; \; ,\; \;  \label{H_redshift_casimir} \\ 
    \displaystyle  \beta &=r_0\left( 1 - \frac{1}{\omega}  \right) + \frac{r_0^2}{\omega r} \; ,  \label{H_shape_casimir}
\end{align}
where $\omega= -r_0^2/\kappa$.
Note that the value of $\omega$ obtained is the same as in the spherical symmetry case, except for the opposite sign; this does not prevent the redshift from being defined over the entire domain of $r$, provided that $\omega$ satisfies:
\begin{align}\label{c0_w_condition}
        \omega < - 1    \; \; \; & \text{or also} & \; \; \; \omega > 0  \; .
\end{align}
In both cases, the redshift is well-defined at the throat. However, considering the result in \eqref{H_redshift_casimir}, we will focus on the case $\omega < -1$. It is worth emphasizing that expressions (\ref{H_redshift_casimir}) and (\ref{H_shape_casimir}) correspond to the metric functions of a traversable wormhole geometry that, to the best of our knowledge, has not been previously reported. It would therefore be interesting to study its behavior and physical implications. Nonetheless, as stated before, in this work we are interested in constructing a traversable wormhole geometry such that the matter source is entirely contained within the region $r < r_{\Sigma}^{e} < 2M$, satisfying the Darmois matching conditions. However, it is straightforward to see that this solution would require the presence of a thin shell as the pressure do not vanish at least $r_\Sigma^{e}$ goes to infinity (which is forbidden in our construction). Now, instead of discarding (\ref{H_redshift_casimir}) and (\ref{H_shape_casimir}) entirely, we shall follow the strategy proposed in \cite{Avalos:2022tqg}, which consists of generalizing the redshift function (\ref{H_redshift_casimir}) and the complexity factor, which in this case reads
\begin{align}\label{H_casimir_complexity_factor}
    Y_{TF} = \displaystyle \frac{r_{0} \left(- 6 r^{2} + 26 r r_{0} - 10 r_{0}^{2}\right)}{3r^{4} \left(- 3 r +  r_{0}\right)}    \;  ,
\end{align}
with the aim to obtain a differential equation for the shape function $\beta$. To be more precise, in this work, we propose the generalized redshift function and the complexity factor to be
\begin{align}
    \alpha_g (r)&=-\frac{1}{2} \log \bigg| \frac{c_0 r}{c_0 r +r_0} \bigg|, \label{generalized_redshift_h}\\
    Y_{gTF} &= \displaystyle \frac{r_{0} \left(a_2 r^{2} + a_1 r r_{0} + a_0 r_{0}^{2}\right)}{r^{4} \left(c_0 r + r_{0}\right)}\nonumber\\ 
    &+ \frac{1}{2r^3} \left( r_0 \beta_{g}'(r_0) - 3\beta_{g}(r_0) \right) \label{GC_complexity}
\end{align}
where \( c_0, \, a_0, \, a_1, \, a_2 \) $\in\mathbb{R}$ are arbitrary constants. If we replace (\ref{generalized_redshift_h}) and
(\ref{GC_complexity}) in (\ref{complex2}) we obtain, after some algebraic manipulation, the following equation for the shape function
\begin{align}
     \frac{d \beta_g}{dr} + \beta_g A(r)   = B(r) \; \;
\end{align}
where
\begin{align}
        A(r) & =   -\frac{c_0}{c_0 r +r_0}-\frac{6}{r}, \nonumber \\ 
         B(r) &  =  -6+\frac{r_0}{c_0 r+r_0}-4a_2-\frac{4a_1 r_0}{r}-\frac{4a_0 r_0^2}{r^2}. \nonumber
\end{align}
From here, it is straightforward to show that the general solution is given by
\begin{align} \label{bebetag}
    \beta_g(r) &= c_1 \frac{r^6(c_0 r +r_0)}{r_0^6}+ \frac{4 a_0 r_0}{7 r^7} - \frac{c_0^5}{r_0^5 (c_0 r + r_0)}\nonumber\\    
    &+r^6 (c_0 r + r_0) \Bigg( 
    -\frac{2 (-a_1 + a_0 c_0)}{3 r^6}\nonumber\\
    &+ \frac{c_0^4 (1 + 4 a_2 - 4 a_1 c_0 + 4 a_0 c_0^2)}{r r_0^5}\nonumber\\
    &- \frac{c_0^3 (1 + 2 a_2 - 2 a_1 c_0 + 2 a_0 c_0^2)}{r^2 r_0^4}\nonumber\\
    &+ \frac{c_0^2 (3 + 4 a_2 - 4 a_1 c_0 + 4 a_0 c_0^2)}{3 r^3 r_0^3}\nonumber\\
    &- \frac{c_0 (1 + a_2 - a_1 c_0 + a_0 c_0^2)}{r^4 r_0^2} \nonumber\\
    &+ \frac{5 + 4 a_2 - 4 a_1 c_0 + 4 a_0 c_0^2}{5 r^5 r_0}
    \nonumber\\
    &+ \frac{4 c_0^5 (a_2 + c_0 (-a_1 + a_0 c_0))}{r_0^6} \log \bigg|\frac{r}{c_0 r +r_0} \bigg| \Bigg), 
\end{align}
where $c_1$ is an integration constant. Note that if we impose the conditions $c_1 =0$ and $a_2= a_1 c_0-a_0 c_0^2$, the expression (\ref{bebetag}) simplifies in the sense that both the $r^{6}$ and the logarithmic terms disappear. These terms could, in principle, be kept, but we have chosen to eliminate them for simplicity and to make the solution more manageable. Now, after explicit replacement, we obtain 
\begin{equation}
   \beta_g(r)= r + \frac{2}{3} c_0 (a_1 - a_0 c_0) r 
+ \frac{2}{21} (7 a_1 - a_0 c_0) r_0 
+ \frac{4 a_0 r_0^2}{7 r}. 
\end{equation}
Finally, using the condition $\beta_g(r_0)=r_0$ we find that $a_1=a_0(c_0-6/7)$, so
\begin{align}
    \beta_g(r)&=r - \frac{4 a_0 c_0 r}{7} - \frac{4 a_0 r_0}{7} + \frac{4 a_0 c_0 r_0}{7} + \frac{4 a_0 r_0^2}{7r}.
\end{align}

Notice that this function does not need to be asymptotically flat since we will impose the coupling conditions with the respective surfaces. We must remember that there are also the mass parameters $M$ and radius $r_\Sigma$, which define the boundary of our object in the sense that it is where our solution lies. In order to make physical sense, the metrics and pressures must be continuous at $r_\Sigma$ where the wormhole and the hyperbolic vacuum meet. The conditions are
\begin{align}
    e^{2\alpha_g(r_\Sigma)}&=\frac{2M}{r_\Sigma}-1, \label{cf}\\
    \frac{\beta_g(r_\Sigma)}{r_\Sigma}-1&= \frac{2M}{r_\Sigma}-1, \label{cf2} \\
    P_r (r_\Sigma)&=0. \label{cf3}
\end{align}
From (\ref{cf2}) and (\ref{cf3}) we obtain
\begin{align}
    c_0 &= -\frac{r_0}{2M}, \\
    a_0 &= -\frac{7M r_\Sigma}{2r_0 (r_\Sigma -r_0)},
\end{align}
which lead to
\begin{align}
    \alpha_g(r)&=-\frac{1}{2}\log \frac{r}{2M-r}=\frac{1}{2} \log \bigg( \frac{2M}{r}-1 \bigg)\label{alphag}\\
    \beta_g(r)&=-\frac{1}{r_\Sigma-r_0}\left(r_{0}r-(2M+r_0)r_\Sigma+\frac{2M r_\Sigma r_0}{r}\right)\label{bebetag2}
\end{align}
It is important to highlight at this point an interesting and useful consequence of the restriction $r \leq \beta_g(r) \leq 2r$ together with the coupling condition (\ref{cf2}), which allows us to derive a constraint on the compactness $M/r_\Sigma$,
\begin{equation}
\frac{1}{2} \leq \frac{M}{r_\Sigma} \leq 1. \label{eqcomparcticitylimit}
\end{equation}

To delimit the possible values of the mass and radius on the external surface, we make use of the condition $r \leq \beta_g(r) \leq 2r$. The first equality, $\beta_g(r) = r$, can be satisfied at two points: $r_1 = r_0$ and $r_2 = 2M$. The first corresponds to the condition at the throat, while the second requires that $r_2 = 2M \leq r_\Sigma$, which is always satisfied under the compacticity restriction. The second equality is satisfied when
\begin{align}
    r_\pm &=\frac{2 M r_{\Sigma} + r_0 r_{\Sigma}}{2 (2 r_{\Sigma}-r_0)}\nonumber\\
    &\pm \frac{\sqrt{r_{\Sigma}} \sqrt{8 M r_0^2 + 4 M^2 r_{\Sigma} - 12 M r_0 r_{\Sigma} + r_0^2 r_{\Sigma}}}
{2 (2 r_{\Sigma}-r_0)}.
\end{align}

To avoid having real roots in the $(r_0, r_\Sigma)$ interval, the discriminant must be less than zero. The limit condition when there is one real root means that the discriminant is zero, i.e.,
\[
8 M r_0^2 + 4 M^2 r_{\Sigma} - 12 M r_0 r_{\Sigma} + r_0^2 r_{\Sigma} \leq 0,
\]
which leads to an upper bound for the total mass:
\begin{equation}
    \frac{1}{2} r_\Sigma \leq M \leq \frac{-2 r_0^2 + 2 \sqrt{r_0^2 (2 r_{\Sigma}-r_0) ( r_{\Sigma}-r_0)} + 3 r_0 r_{\Sigma}}{2 r_{\Sigma}}.
\end{equation}

Thus, for each value of $r_\Sigma$, there exists a minimum and a maximum allowed mass:
\begin{align}
M_{\text{min}}(r_\Sigma) &= \frac{r_\Sigma}{2}, \\ 
M_{\text{max}}(r_\Sigma) &= \frac{-2 r_0^2 + 2 \sqrt{r_0^2 (2 r_{\Sigma}-r_0) ( r_{\Sigma}-r_0)} + 3 r_0 r_{\Sigma}}{2 r_{\Sigma}}.
\end{align}

\begin{figure}[t!]
    \centering
    \includegraphics[scale=1]{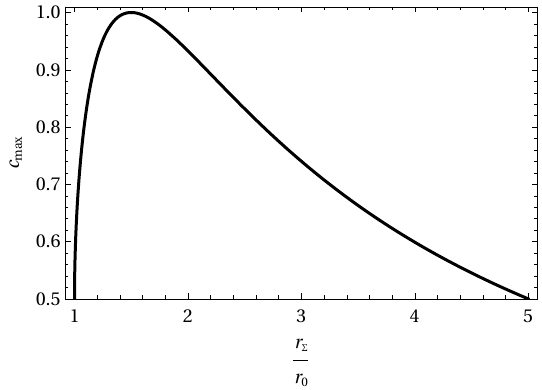}
    \caption{Maximum compacticity $c_{\text{max}}(r_\Sigma)$ as a function of the object's radius $r_\Sigma$. The maximum campacticity $c_{max}=1$ is reached when the radius is $r_\Sigma=3r_0/2$.}
    \label{fig:compacticity}
\end{figure}

Accordingly, the compacticity is constrained by
\begin{align}
    \frac{1}{2} \leq \frac{M}{r_\Sigma} \leq& \frac{3r_0}{2 r_\Sigma} - \frac{r_0^2}{r_\Sigma^2} + \frac{r_0^2}{r_\Sigma^2} \sqrt{\left(2\frac{r_\Sigma}{r_0} - 1\right) \left(\frac{r_\Sigma}{r_0} - 1\right)}\nonumber\\
    &= c_{\text{max}}(r_\Sigma).
\end{align}

The maximum compacticity $c_{\text{max}} = 1$ is attained at $r_{\Sigma} = \frac{3}{2}r_0$, while the minimum value $c_{\text{min}} = \frac{1}{2}$ occurs at $r_\Sigma = r_0$ and $r_\Sigma = 5r_0$, as shown in Fig.~\ref{fig:compacticity}. For the wormhole solution to be physically acceptable, the parameters $r_\Sigma$ and $M$ must satisfy the following conditions:
\begin{align}
    r_0 \leq &r_{\Sigma}\leq 5r_0, \label{valoresr} \\
    \frac{r_{\Sigma}}{2} \leq &M \leq \frac{3r_0}{2 } - \frac{r_0^2}{r_\Sigma} + \frac{r_0^2}{r_\Sigma} \sqrt{\left(2\frac{r_\Sigma}{r_0} - 1\right) \left(\frac{r_\Sigma}{r_0} - 1\right)}. \label{valoresm}
\end{align}

Finally, the flaring-out condition is always fulfilled, since
\begin{equation}
    \beta_g '(r_0) = \frac{2Mr_\Sigma - r_0^2}{r_0 r_\Sigma - r_0^2} > 0,
\end{equation}
for all allowed values of the mass $M$. Now, using (\ref{alphag}) and (\ref{bebetag}), the energy density and radial pressure become,
\begin{align}
\rho (r)&= -\frac{2M r_\Sigma-r^2 }{r_\Sigma -r_0} \frac{r_0}{8\pi r^4}, \\
    P_r(r)&= \frac{r_\Sigma -r}{r_\Sigma-r_0} \frac{r_0}{8 \pi r^3},
\end{align}
from where
\begin{equation}
        \rho (r)+P_r(r)= -\frac{2M -r }{r_\Sigma -r_0} \frac{r_0 r_\Sigma}{8\pi r^4}<0,
\end{equation}
so that the solution violates the null energy condition everywhere, from the radius of the throat $r_0$  to the radius of the entire object $r_\Sigma$. At this point, it is worth highlighting an interesting feature not present in previous studies on Casimir wormholes. While both \cite{lobo} and \cite{Avalos:2022tqg} claim that their solutions require only an arbitrarily small amount of exotic matter, in those cases, the exotic matter is spread throughout the entire spacetime. In contrast, in our case, we find that the exotic matter can be confined to an arbitrarily small region of the spacetime without requiring thin shells of matter.\\

\begin{figure}[t!] 
    \centering
    \includegraphics[width=0.45\textwidth]{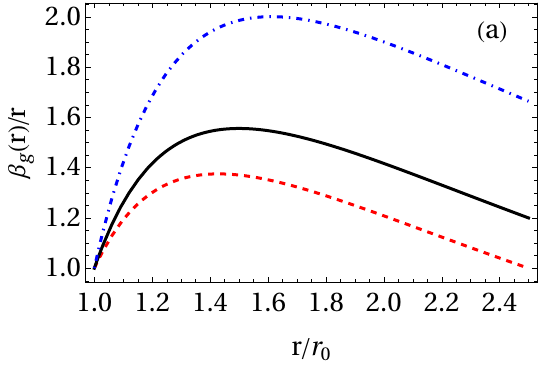}\
    \includegraphics[width=0.45\textwidth]{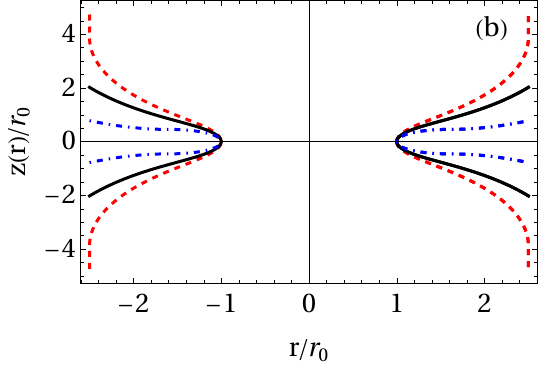}\
    \includegraphics[width=0.5\textwidth]{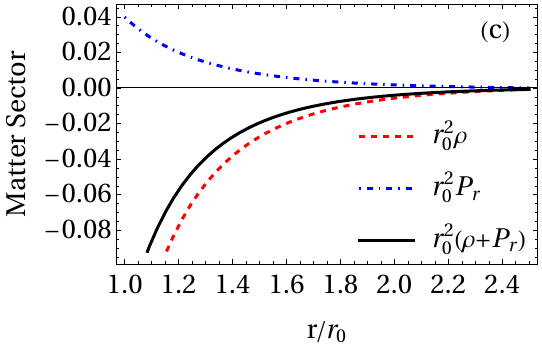}
    \caption{(a) $\beta_g/r$ versus $r/r_0$ for $r_{\Sigma} = 2.5r_0$ and various values of $M$: $M = M_{\text{min}}(r_{\Sigma}) = 1.25 r_0$ (red dashed), $M = 2.5r_0$ (black solid), and $M = M_{\text{max}}(r_{\Sigma}) = 2.0798r_0$ (blue dot-dashed). (b) Embedding diagram: $z/r_0$ as a function of $r/r_0$ for the same values of $r_{\Sigma}$ and $M$ as in (a). (c) Normalized energy density $r_0^2 \rho$ (red dashed), normalized radial pressure $r_0^2 p_r$ (blue dot-dashed), and their sum $r_0^2(\rho + p_r)$ (black solid) versus $r/r_0$, for fixed $r_{\Sigma} = 2.5r_0$ and $M = 1.5r_0$.}
    \label{fig:solution}
\end{figure}

To complement the discussion on the sector supporting the wormhole geometry, in Fig.~\ref{fig:solution} we display the profiles of the shape function, the embedding diagram, and the matter sector: energy density $\rho$, radial pressure $p_r$, and the combination $\rho + p_r$ for the parameter values indicated in the legend. We omit the plot of $\rho + p_t$, as this quantity remains positive throughout the spacetime.
Panel (a) of Fig.~\ref{fig:solution} shows $\beta_g/r$ as a function of the radial coordinate for the chosen parameters. Note that the speed of convergence increases as the total mass decreases. In particular, the minimal mass configuration (red dashed line) reaches the lower bound $\beta_g = r$ exactly at the boundary $r = r_\Sigma$, while the maximum mass configuration (blue dot-dashed line) reaches the upper bound $\beta_g = 2r$ at an intermediate value of $r$. For intermediate values of the mass, neither limit is reached within the domain. Panel (b) displays the embedding diagram corresponding to the same parameter choices. As $r$ increases, the derivative $dz/dr \to 0$ more rapidly for higher total masses. In contrast, for the minimum mass configuration, the derivative diverges at $r = r_\Sigma$, as the shape function satisfies $\beta_g(r_\Sigma) = 2r_\Sigma$. Panel (c) presents the matter sector. Both the energy density $\rho$ and the radial pressure $p_r$ exhibit a qualitatively similar behavior to that reported in \cite{garattini2019casimir}, in the sense that they are negative throughout the radial domain, reminiscent of the Casimir-like profile. However, this similarity is merely formal: in our case, the equation of state is determined by Eq.~(\ref{EoS_hyperbolic}), and the behavior ultimately depends on the free parameters $M$ and $r_\Sigma$. While the energy density remains negative for all $r$ and asymptotically approaches positive values, the Casimir-like behavior breaks down for the radial pressure.\\ 

\begin{figure*}[htb!] 
    \centering
    \includegraphics[width=0.45\textwidth]{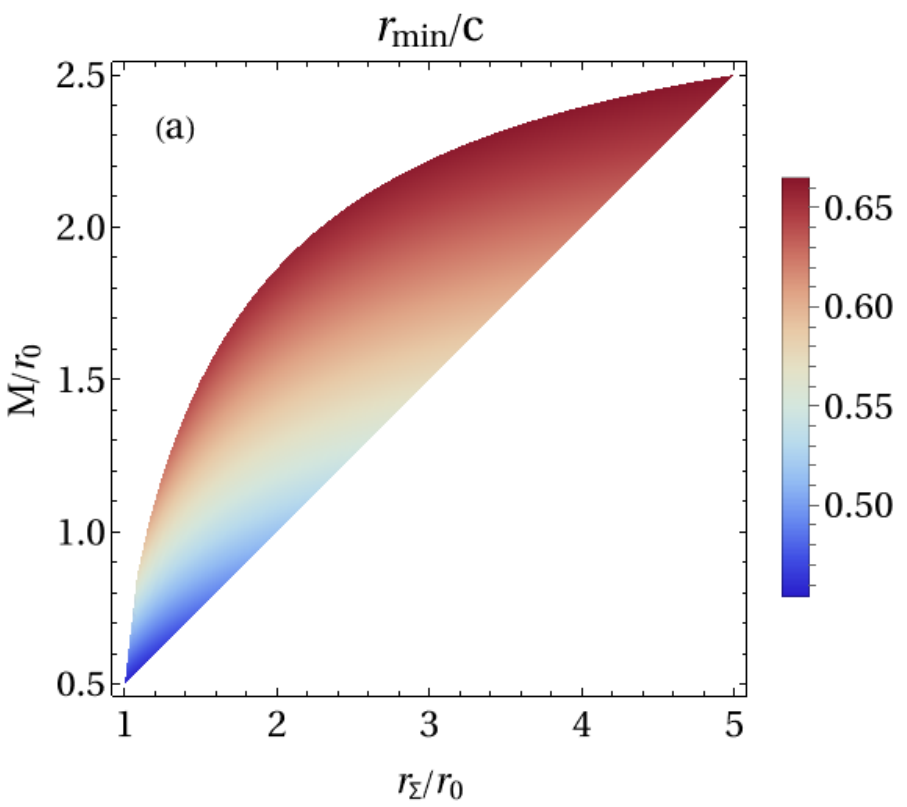}\
    \includegraphics[width=0.45\textwidth]{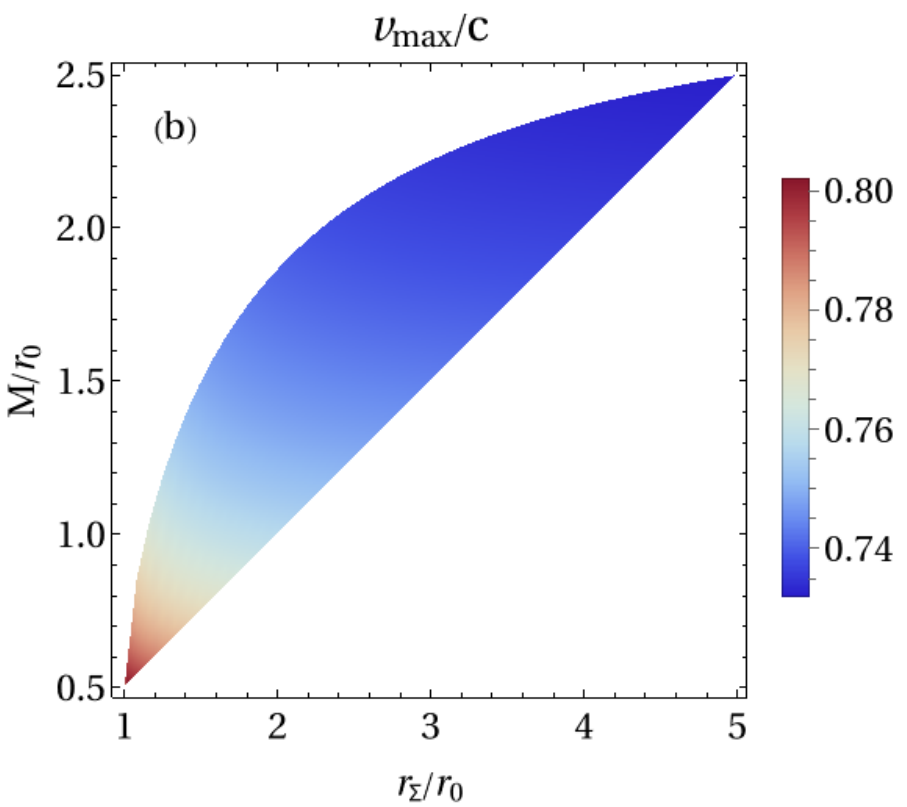}\
    \includegraphics[width=0.45\textwidth]{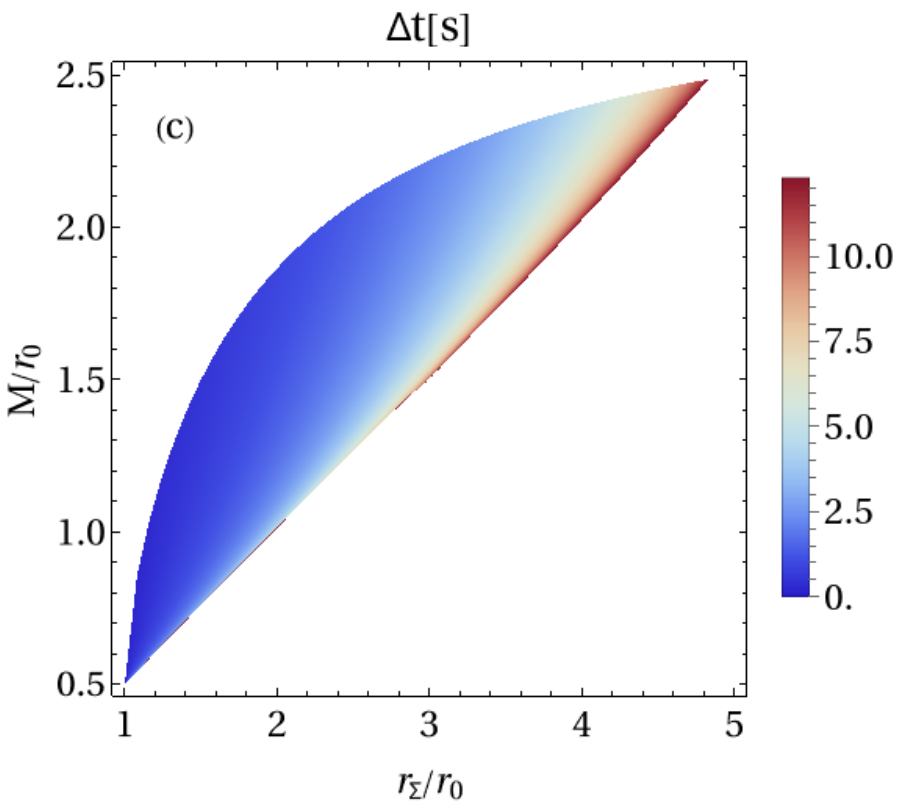}\
    \includegraphics[width=0.45\textwidth]{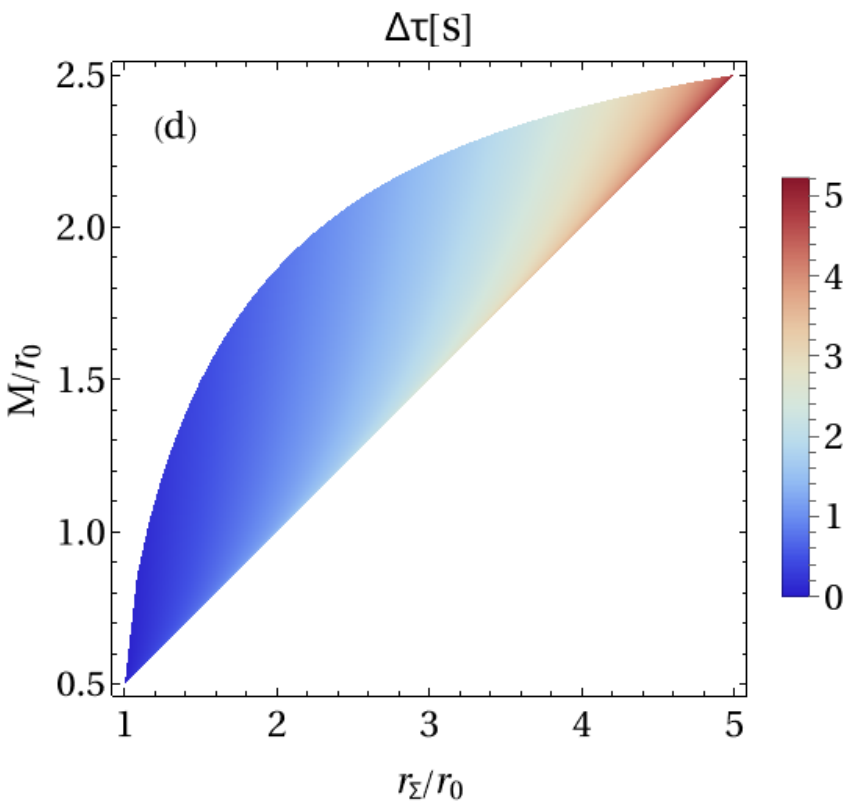}
    \caption{Minimum throat radius $r_{min}$ (panel (a)) and maximum traveler velocity $v_{max}$ (panel (b)) that fulfill humanly traversable conditions for all possible values of normalized mass $M/r_0$ and normalized cavity radius $r_\Sigma/r_0$. For the same mass and radius parameters the corresponding coordinate time for crossing the wormhole $\Delta t$ (panel (c)) and a proper time $\Delta \tau$ (panel (d)).}
    \label{fig:marea}
\end{figure*}

Since our wormhole satisfies the basic requirement of being traversable, it is natural to explore whether this geometry could, in principle, allow for human interstellar travel. To that end, we must ensure that the tidal accelerations experienced by a traveler remain of the same order or less than the gravitational acceleration on Earth. The requirement arises from the necessity to ensuring the physical plausibility and biological survivability of the traveler. These tidal accelerations arise as a direct consequence of spacetime curvature: in general relativity, the relative acceleration between neighboring free-falling particles is governed by the geodesic deviation equation, which involves contractions of the Riemann curvature tensor. Physically, this means that different parts of an extended body, such as a person, will follow slightly different geodesics due to the non-uniformity of the gravitational field. The resulting differential acceleration is what we perceive as a tidal force. In the context of wormholes, particularly near the throat where curvature effects are most pronounced, these tidal effects can become dangerously large unless the geometry is carefully controlled.  In particular, the radial and lateral tidal accelerations must satisfy the following constraints:
\begin{align}
&\left| \left( \frac{\beta_g}{r} -1\right) \left[  \alpha_g'' + (\alpha_g ')^2 -\frac{r \beta_g' -\beta_g}{2r(r-\beta_g)} \alpha_g ' \right] \right| |\eta^1| \le \frac{g_{\oplus}}{c^{2}}, \label{tidalr} \\
&\left| \frac{\gamma^2}{2r^{2}}\left[\frac{v^{2}}{c^2}\left(\beta_g' - \frac{\beta_g}{r}\right) + 2r(r-\beta_g)\alpha_g'\right] \right| |\eta^{2}| \le \frac{g_{\oplus}}{c^{2}}, \label{tidalt}
\end{align}
where $g_{\oplus}$ is the gravitational acceleration on Earth, $\gamma = 1/\sqrt{1 - v^2/c^2}$, $c$ is the speed of light, and $\eta^1$, $\eta^2$ denote the radial and lateral sizes of the traveler, respectively. We shall use Eqs.~(\ref{tidalr}) and (\ref{tidalt}) to estimate the size of the throat and the velocity of the traveler at the throat (assumed constant for simplicity). At the throat, the constraints reduce to:
\begin{align}
&|\alpha'_g(r_0)| \le \frac{2g_{\oplus}r_0}{(\beta_g'(r_0) - 1)|\eta^1| c^2}, \label{rmin} \\
&\gamma^2 v^2 \le \frac{2g_{\oplus}r_0^2}{(\beta_g'(r_0) - 1)|\eta^2|}. \label{vmax}
\end{align}
Saturating inequalities (\ref{rmin}) and (\ref{vmax}) defines a minimum throat radius $r_{\text{min}}(r_\Sigma, M)$ and a maximum velocity $v_{\text{max}}(r_\Sigma, M)$. In addition, to ensure the trip is completed in a reasonable amount of time, we require both the coordinate and proper times to be bounded as follows:
\begin{align}
\Delta t &= \int\limits_{r_{0}}^{r_{st}} \frac{e^{-\alpha_{g}} \, dr}{v \sqrt{\beta_{g}/r - 1}} < 1\ \text{year}, \label{coort} \\
\Delta \tau &= \int\limits_{r_{0}}^{r_{st}} \frac{dr}{v \gamma \sqrt{\beta_{g}/r - 1}} < 1\ \text{year}, \label{propt}
\end{align}
where $r_{st}$ denotes the radial coordinate of the station. Although exact expressions for $r_{\text{min}}(r_\Sigma, M)$ and $v_{\text{max}}(r_\Sigma, M)$ can be derived, the numerical solution provides clearer insight into the behavior of the system. For our analysis, we take $\eta^1 \approx 2\,\text{m}$ and $\eta^2 \approx 1\,\text{m}$ as typical human dimensions, and set $r_{st} = r_\Sigma$, representing a journey from the outer surface to the throat of the wormhole. The numerical results are shown in Fig.~\ref{fig:marea}. An interesting feature is that the coordinate time $\Delta t$ diverges when the compacticity reaches $M/r_\Sigma = 1/2$, as the integrand behaves like
\begin{equation}
    \frac{e^{-\alpha_{g}}}{\sqrt{\beta_{g}/r - 1}} = \sqrt{\frac{r^3(r_\Sigma - r_0)}{(r - r_0)(r_\Sigma - r_0)^2 r_\Sigma}},
\end{equation}
whose integral diverges. This behavior is expected and mirrors the divergence of coordinate time for an object falling into a black hole when compacticity approaches $0.5$. However, the proper time $\Delta \tau$ remains finite. The numerical analysis shows that the relevant quantities are bounded as follows:
\begin{align}
    0.453703 \le &\frac{r_{\text{min}}}{c}\le 0.665124, \\
    1.35657 \times 10^8\ \text{m} \le r_{\text{min}}& \le 1.98872 \times 10^8\ \text{m}, \\
    0.731865 \le &\frac{v_{\text{max}}}{c}\le 0.804181, \\
    0 \le &\Delta t \le \infty, \\
    0\ \text{s} \le &\Delta \tau\le 5.22062\ \text{s},
\end{align}
where the extremal values are reached at $\{ r_\Sigma = r_0,\, M = r_\Sigma / 2 = r_0 / 2 \}$ and $\{ r_\Sigma = 5r_0,\, M = 3r_0 / 2 \}$, corresponding to the boundaries of the allowed region.
Note that both the proper and coordinate times vanish when $r_\Sigma = r_0$, since the traveler would be moving directly from the hyperbolic vacuum to the throat. Another remarkable observation is that the minimum radius of a humanly traversable wormhole throat is on the same order of magnitude as the Earth–Moon distance which is approximately 384,400 km. While this distance varies throughout the year due to the Moon's elliptical orbit, the average value serves as a useful reference scale for comparison.

\section{Final remarks}\label{concl}

We have studied an interior solution for a hyperbolic wormhole which, to the best of our knowledge, is the first time such an analysis has been carried out, although the idea of a wormhole with this symmetry has previously been proposed in a different context \cite{lobo2010possibility}. Our study provides a complete framework for constructing hyperbolic wormhole geometries belonging to the same ``equivalence class" as the Casimir wormhole reported in \cite{garattini2019casimir, Avalos:2022tqg}. The source of the wormhole is modeled by hyperbolically symmetric static fluid distributions that are locally anisotropic.\\

To obtain the solution, we implemented the recently introduced concept of complexity \cite{Complex}, which has been widely used as a supplementary condition to close the Einstein field equations in various scenarios \cite{Annals2, IJMP, Yousaf, Abbas, Abbas2, Herrera3, Sharif:2021, 2Sharif:2021, Annals}.  The solution is fully characterized by two parameters, $r_\Sigma$ and $M$, which arise from a generalization of the redshift function used in the Casimir wormhole \cite{garattini2019casimir}. These parameters are constrained by imposing fundamental requirements for traversability: (i) the flaring-out condition, (ii) tidal accelerations comparable to or smaller than Earth's surface gravity, and (iii) finite travel time from a spatial station to the throat. Perhaps the most striking feature of the solution is that it corresponds to an interior geometry located within the event horizon. Unlike most traversable wormhole solutions, which are asymptotically flat, our model requires the application of Darmois junction conditions to match the wormhole geometry \eqref{lobo_metric} with the hyperbolic vacuum solution \eqref{hyperbolyc_empty_metric}, as opposed to matching with a flat or de Sitter exterior \cite{garattini2019casimir,Avalos:2022tqg, Bhar}. It is worth mentioning that, in contrast to what generally occurs, the solution satisfies the Darmois condition without requiring any thin shell.\\

It is well known that minimizing exotic matter is a major challenge in wormhole physics. Several methods have been proposed to reduce its presence, including the cut-and-paste approach \cite{Visser1, Visser2}, and the volume integral quantifier introduced by Visser \cite{Visser3}, which estimates the total amount of exotic matter required to keep the wormhole throat open. In contrast to what is typically found in spherical symmetry, hyperbolic symmetry naturally gives rise to negative energy densities so the notion of ``minimizing exotic matter" becomes unnecessary, representing a distinct advantage of the approach. In this regard, our study demonstrates that alternative geometries can provide a consistent framework to overcome some of the limitations of general relativity, opening new avenues for understanding gravitational phenomena on both astrophysical and cosmological scales. From an astrophysical standpoint, while the exact realization of hyperbolic wormhole geometries remains speculative, their study may inform effective models of high-curvature regions such as the interiors of compact objects or early-universe phases. Moreover, the natural emergence of negative energy densities in hyperbolic symmetry could have implications for semiclassical gravity or quantum field theory in curved spacetime, where such effects are expected to arise. These considerations provide a physical motivation for exploring geometries beyond the usual spherical symmetry.
 \\

This work has revealed new possibilities for constructing hyperbolic wormhole geometries and their potential application in modeling other astrophysical phenomena, laying the groundwork for future research. Extensions of this analysis could include scalar potentials, possibly yielding new stable configurations. Other promising directions involves rotating wormholes with cosmic strings, or wormholes within the framework of alternative theories of gravity \cite{Mazur:2001fv, KG,  R}. In particular, a detailed stability analysis is crucial to assess the physical viability of these solutions, though such an analysis lies beyond the scope of the present work and will be addressed in future studies. \\

\section{Acknowledgements}
 E. F. is grateful for / MINCYT-CDCH-UCV/ 2024 - and acknowledges support from Consejo de Desarrollo Cientfico y Humanstico - Universidad Central de Venezuela in part by a grant entitled Study of compact stellar configurations composed of spherically symmetric and hyperbolic, static and anisotropic relativistic fluids in the context of General Relativity. Also, E. F.  is grateful for the Funcaci\'on Carolina-$25$ years for granting the short-stay scholarship that served to finance this work.  E. C. is funded by the Beatriz Galindo contract BG23/00163 (Spain). E. C. acknowledge Generalitat Valenciana through PROMETEO PROJECT CIPROM/2022/13.


\bibliographystyle{unsrt}
\bibliography{references}

\end{document}